\documentclass[superscriptaddress, aps,pre, twocolumn, floatfix]{revtex4-2}
\usepackage{graphicx}
\usepackage{dcolumn}
\usepackage{amsmath}
\usepackage{latexsym}
\usepackage{mathrsfs}
\usepackage{mathtools}
\usepackage[normalem]{ulem} 
\usepackage[usenames,dvipsnames]{color}
\usepackage{lineno}

\usepackage[breaklinks,colorlinks = true,linkcolor = blue,urlcolor=blue,citecolor=blue]{hyperref}

\hypersetup{
    colorlinks = true,
    linkcolor =blue,
	citecolor=blue, 
	urlcolor=blue 
}

\newcommand{\la}{\langle}
\newcommand{\ra}{\rangle}

\newcommand{\ed}{\end{document}}
\newcommand{\beq}{\begin{equation}}
\newcommand{\eeq}{\end{equation}}

\usepackage{mathtools}
\usepackage{comment}

\begin{document}


\title{Hydrodynamic description of correlation matrix eigenvalues in a finite temperature expanding free fermionic gas }

\author{Atharva Naik}
\email{atharva.naik@icts.res.in} 
\affiliation{International Centre for Theoretical Sciences, Tata Institute of Fundamental Research,
Bangalore 560089, India}

\author{Madhumita Saha}
\email{madhumita.saha@icts.res.in }
\affiliation{International Centre for Theoretical Sciences, Tata Institute of Fundamental Research,
Bangalore 560089, India}

\author{David A. Huse}
\email{huse@princeton.edu } 
\affiliation{Department of Physics, Princeton University, NJ 08544, USA}

\author{Manas Kulkarni}
\email{manas.kulkarni@icts.res.in }
\affiliation{International Centre for Theoretical Sciences, Tata Institute of Fundamental Research,
Bangalore 560089, India}
\author{Abhishek Dhar}
\email{abhishek.dhar@icts.res.in }
\affiliation{International Centre for Theoretical Sciences, Tata Institute of Fundamental Research,
Bangalore 560089, India}

\begin{abstract}
A recent work considered the Page curve for entanglement entropy of a freely expanding one-dimensional Fermi gas starting from a finite box (the system) in a zero-temperature initial state that then expands in to a semi-infinite vacuum (the environment).  For the case with a single defect at the system-environment junction, it was shown that the Yang-Yang entropy of the system computed from semiclassical hydrodynamics agrees with the microscopically computed entanglement entropy, yielding the Page curve. On the other hand, the environment Yang-Yang entropy showed a monotonic increase.  Here, we propose a hydrodynamic description to obtain the eigenvalues of the microscopic correlation matrix for a nonzero temperature freely-expanding Fermi gas. For the system, the  phase space density directly gives the correlation matrix eigenvalues and thus, via the Yang-Yang formula, yields the von-Neumann entropy. 
For the reservoir, a modification of the phase space density is required in the Yang-Yang formula, to incorporate coherences, in order to reproduce accurately the entropy. The hydrodynamic description of correlation matrix eigenvalues also leads to a  Levitov-Lesovik type formula for the full-counting-statistics for a finite system in the transient regime. We provide a heuristic derivation of the correlation matrix eigenvalues based on a semiclassical wave-packet picture.

\end{abstract}
\date{\today}

\maketitle
{\it Introduction---} 
 Understanding the dynamics and steady states of quantum systems  is one of the interesting fundamental problems, one goal being to find the truly quantum effects~\cite{BP_book,Daley_review,amrey,CPPG2021}.  The study of many-body systems with a large number of particles at long times is challenging, both analytically and numerically and  several efforts are going on to address this problem, even at the level of non-interacting particles and interacting integrable systems~\cite{cef2011,SFRKL2011,Eisler_2012,ef2016,rigol2007,vidmar2016,Caux_2016,DSVC2017,FC2010,SKA2021,GLC2020}. 
One of the theoretical approaches to understand  nonequilibrium dynamics  of such systems is through hydrodynamics. For integrable systems, generalized hydrodynamics~\cite{BFPC2018,ghd_prl,ghd_prx,bd2020} captures many of the essential features of the  dynamics, including density profiles, currents, number fluctuations and entanglement.
 
In the context of nonequilibrium dynamics, one paradigmatic nonequilibrium initial condition is the
so-called domain wall state~\cite{antal,gd_pre}. For the case of non-interacting fermions this has been studied widely
and several important and illumunating results have been obtained. Two important classes of results are on the
evolution of entanglement and of number fluctuations. Specifically consider an infinite one-dimensional lattice with a defect at the origin, with the left half, 
$x < 0$ (called the system, S), initially occupied and all other sites (reservoir, R) initially empty. As the domain wall melts, particles escape at a constant rate from the
system to the reservoir and entanglement between the system and reservoir grows linearly with time. The fluctuations
in the escaped particle number also grow linearly with time as do all cumulants, which are given completely
by the Levitov-Lesovik formula for the cumulant generating function~\cite{Levitov1993,LL2,KL09,FS2025} $\mu(\lambda)=t \int_{0}^\pi dk v_k/2 \pi [1+e^{(\lambda-1)}T_k]$, where $v_k$ and $T_k$ are the group velocity and transmission probability of particles at wavevector $k$. Another
remarkable result is that of Klich and Levitov, which provides an exact relation between the entanglement
entropy (EE) and the full counting statistics (FCS) for number flutuations. The exact computations of EE and $\mu(\lambda)$ are highly nontrivial but it turns
out that they can be obtained from the semiclassical phase space dynamics of the particles. In particular, for the infinite domain-wall,
the so-called Yang-Yang (YY) entropy, written in terms of the phase space density, $n(x, k, t)$, as
\begin{eqnarray}
\label{eq:YY}
S_{YY} &=& \int_{D} dx \int \frac{dk}{2\pi} \bigg[ n(x,k,t)\log n(x,k,t) \nonumber \\ &+& 
\left(1 - n(x,k,t)\right)
\log\left(1 - n(x,k,t)\right) \bigg]
\end{eqnarray}
is known~\cite{alba2017,alba2021} to capture accurately the entanglement entropy of the system (with domain of integration $D = S$) and of the reservoir (for $D = R$). The fact that the YY entropy captures the EE was conjectured in Ref~\cite{SciPostPhys_Conj}  but, to our knowledge, a microscopic derivation is so far lacking. One of the main goals in this Letter is to provide a semiclassical derivation.
 
Next, consider the case of the domain wall set-up but with  a finite lattice system  with $L_S$ sites. This corresponds to the physical situation of free expansion of particles from a finite box into the vacuum.  This has recently been studied as a simple toy model which exhibits the dynamic Page curve~\cite{SKD2024,Liu2021,Kehrein2024,Glatthard2024,LKG2025,JMK2025,RDK2025} and the precise form of the entanglement curve can be understood completely using hydrodynamics and the YY formula applied to the system~\cite{SKD2024}. However, it was observed that the YY entropy fails to capture the reservoir EE and this has been an  outstanding problem. Using our semiclassical formulation, we are able to resolve this puzzle and provide a modified YY formula for the reservoir. This is our second important result. 

We consider the setup of a noninteracting spinless fermion gas at a nonzero temperature, initially in a finite box, and which escapes freely into a semi-infinite empty reservoir (the environment), in the presence of a point defect at the system-reservoir junction.  We use a semiclassical wavepacket picture to compute explicitly the correlation matrix eigenvalues and then relate them to  phase space  hydrodynamics of the Fermi gas.  This enables us to analytically predict the evolution of quantities such as von Neumann (vN) entropy and FCS. In particular, we are able to demonstrate the agreement of the microscopically computed entropy with that predicted from a modified version of the  YY entropy, both for the system and the reservoir. 

{\it Model---} We consider a one-dimensional free fermionic chain, consisting of $L_S$ system sites and $L_R$ reservoir sites, with the Hamiltonian given by,
\begin{align}
\label{set-up}
    \hat{H} &= \hat{H}_S + \hat{H}_R + \hat{V},~{\rm where}~~\\
     \hat{H}_S&=-\!\!\!\!\!\!\!\sum_{i = -L_S+1}^{-1} \!\!\!g(\hat{c}^{\dagger}_i \hat{c}_{i+1} + \textrm{h.c}), \hat{H}_R=-\!\!\!\sum_{i = 1}^{L_R-1} g(\hat{c}^{\dagger}_i \hat{c}_{i+1} + \textrm{h.c}),\notag
\end{align}
where $\hat{c}_i\, (\hat{c}^\dagger_i)$ are spinless fermionic annihilation (creation) operators at site $i$, $\hat{H}_{S(R)}$ is the system (reservoir) Hamiltonian and $\hat{V} = -g_c(\hat{c}^\dagger_{0}\hat{c}_1 + \textrm{h.c}) + h_0 (\hat{c}^\dagger_0\hat{c}_0 - \hat{c}^\dagger_1\hat{c}_1)$  is the defect Hamiltonian. A special case is the conformal defect, with the  onsite potential set as $h_0= \sqrt{g^2 - g_c^2}$.   As $L_S$ and $L_R$ approach infinity, the associated scattering problem for the given defect yields transmission probability $T_k $ and reflection probability $R_k =1-T_k$, where $k$ is the wave-vector. For the conformal defect, $R_k \coloneq R= 1 - (g_c/g)^2$  is $k$-independent.  Henceforth we set $g=1/2$ and choose $ 0<g_c<g$.    

{\it  Microscopic approaches and initial conditions ---}
We write the Hamiltonian in Eq.~\eqref{set-up} as,
$\hat{H}=\sum_{i,j=-L_S+1}^{L_R} h_{ij} \hat{c}^{\dagger}_i \hat{c}_j$, which defines the matrix $h$.   The correlation matrix $C(t)$ for any time $t$ is defined as, $C_{ij}(t)=\langle \hat{c}^{\dagger}_i(t) \hat{c}_j(t) \rangle$, where  $\langle...\rangle$ denotes an average over the initial density matrix $\rho(0)$.  The evolution of the correlation matrix is given by, 
 \begin{equation}
     C(t)=e^{i h t} C(0) e^{-i h t}\,.
     \label{eq:Ct}
 \end{equation}
Initially the reservoir is empty, i.e $[C(0)]_{i,j = 1}^{L_R} = 0$, while the system is in a thermal state $\rho_S^{(\beta,\mu)}(0)={e^{-\beta(\hat{H}_S-\mu \hat{N}_S)}}/{Z}$, where $\hat N_S=\sum_{i=-L_S+1}^0 \hat c_i^\dagger \hat c_i$ and $Z$ is the partition function. To obtain the corresponding initial system correlation matrix, we 
 write  the system Hamiltonian in Eq.~\eqref{set-up} as, $
\hat{H}_S\!=\!\!\!\sum\limits_{i,j=-L_S+1}^0 \!\!\!h_{ij}^S \hat{c}_i^{\dagger} \hat{c}_j$.  One then gets (see Supplementary Material~\cite{supp}),
\begin{align}
\label{eq:FT_ini_state}
\langle \hat{c}^{\dagger}_i (0) \hat{c}_j (0) \rangle = \sum\limits_{q=1}^{N} U_{qi}^* U_{jq} f(\lambda_q),
\end{align}
where $U_{jq}$ and  $\lambda^S_q=-\cos q$, with $q=s\pi/(L_S+1),~s=1,2,\ldots,L_S$, denote  the eigenvectors and eigenvalues of the single particle Hamiltonian $h^S$, and $f(\omega)=[1+e^{\beta(\omega-\mu)}]^{-1}$ is the Fermi distribution function. Then $C(t)$ is obtained from Eq.~\eqref{eq:Ct}. The vN entropy and the FCS for the system and reservoir parts can now be obtained from the eigenvalues of the restricted correlation matrices,  which are respectively, $C_{i,j}^S(t)\equiv [C(t)]_{i,j = -L_S+1}^{0}$ for the system part, and $C_{i,j}^R(t)\equiv [C(t)]_{i,j = 1}^{L_R}$ for the reservoir part. Let us denote the non-zero eigenvalues of these correlation matrices by $\zeta_i^D$, $i=1,2,\ldots,L_S$ and with $D=S(R)$ for system (reservoir) parts. The 
 vN entropy and the 
cumulant generating function $\mu(\lambda)=\log \la e^{\lambda \hat N_D}\ra$ for the FCS can be expressed in terms of these eigenvalues as,
\begin{align}
\label{eq:vN_entropy} 
&S_{v}^{D}=-\sum_{l = 1}^{L_S} \left[(1-\zeta^{D}_l) \log[1-\zeta_l^{D}]+ \zeta_l^D \log \zeta_l^D \right],
\\   \label{eq: FCS} &\mu^D(\lambda) = \sum_{l = 1}^{L_S}  \log \Big{[}1 + (e^{\lambda} - 1)\zeta^{D}_l\Big{]}.     
\end{align}
{\it Semiclassical theory:} We now present a semiclassical theory to compute the eigenvalues. 
We begin with an observation made  in Ref.~\cite{NAK2026} (see also~\cite{EP2012}), for a similar setup with a conformal defect (hence $R_k=R$), at $\beta = \infty$  and with an initially fully occupied system. 
At any given time $2n L_{S} < t<2(n+1) L_{S}$, 
with $n=0,1,2,\ldots$, almost all of the eigenvalues of $C_{i,j}^S(t)$ are very accurately given  by  values from the set $\{1, R, R^2, \cdots,R^{n+1}\}$. 
Similarly,  the eigenvalues of $C_{i,j}^R(t)$  are accurately given by values from the set  $\{0, (1-R), (1-R^2),\cdots, (1-R^{n+1})\}$. A small number, $o(L_S)$, of eigenvalues have intermediate values. We discuss now the semiclassical theory, based on a wavepacket description, that provides an understanding of this behavior of the correlation matrix eigenvalues, and generalizes the result to nonzero temperatures.  Following discussion holds for a more general defect model, although we will present numerical results only for the conformal case. 

\begin{figure}[t]
    \centering
    \includegraphics[width=\linewidth]{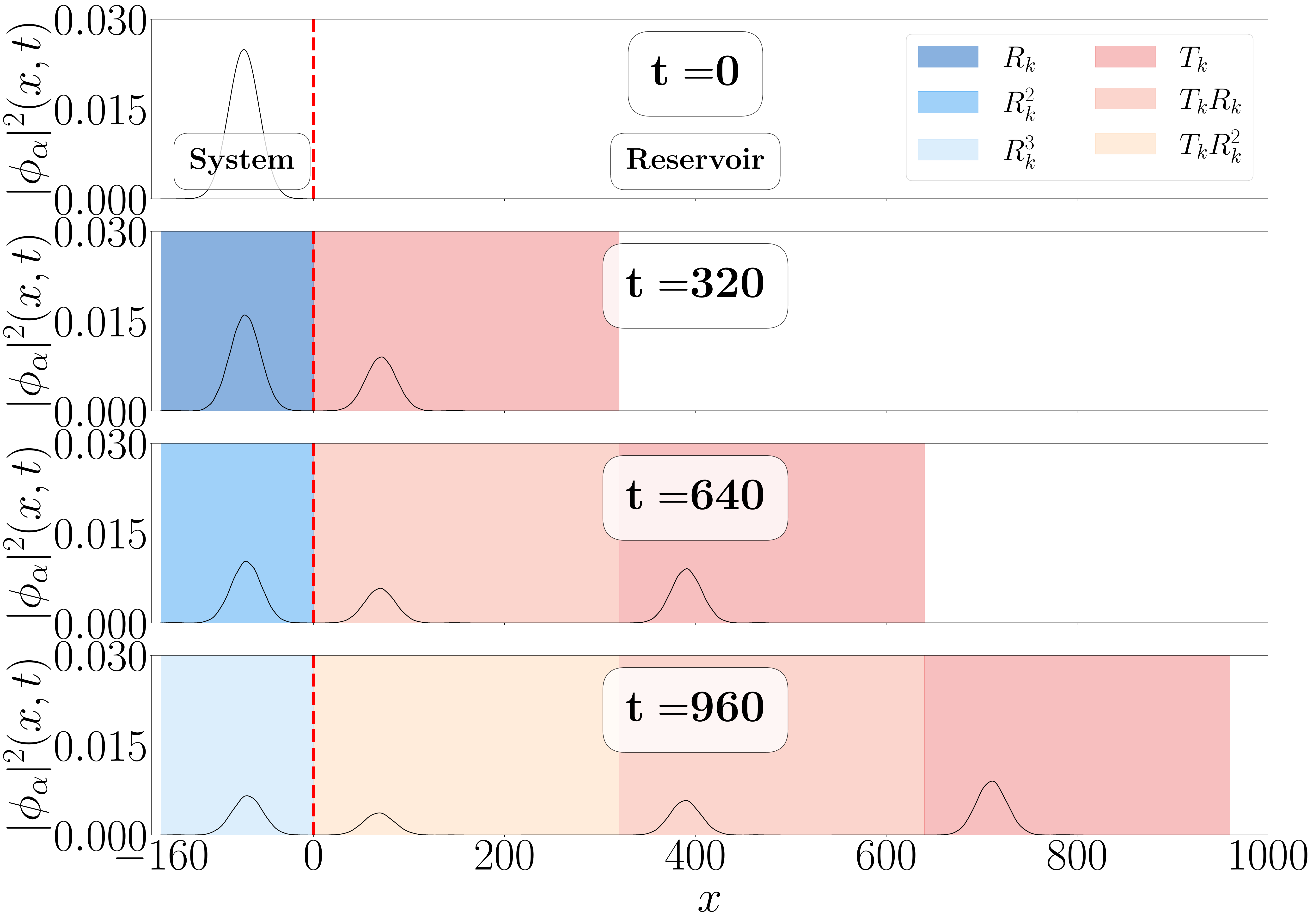}
    \caption{
    $|\phi_{{\alpha}}(x,t)|^2$ vs $x$ for $\alpha = (-80,-\frac{\pi}{2})$ at $t = 0$. Position of the wave packet at times $t = 0$, $320$, $640$, and $960$ is shown. Here, the system size $L_S = 160$ and $g_c = 0.4$. We observe a single bump in the system and a train of bumps in the reservoir.  The varying shades of blue indicate the decay of the amplitude of the quantity $\sum_{-2L_S}^0|\phi_{{\alpha}}(x,t)|^2$ in the system as the time evolves from $R_k$ to $R_k^2$ and then to $R_k^3$. In the reservoir part, varying shades of pink represent different cells corresponding to trains of transmitted packets weighted by $T_k R_k^s$. }
    \label{fig:wavepacket}
\end{figure}

To proceed, we first write the initial state using a different single-particle basis. Consider a basis set with states $|\phi(\alpha)\rangle$, with the index  $\alpha=(y,p)$  indicating states localized around the point $(y,p)$ in phase space, $x \in (-L_S,0),~k \in (-\pi,\pi)$, with spreads $\Delta x =L_S/K, \Delta k=2\pi/\Delta x=2\pi K/L_S$. For an explicit construction of such an orthonormal basis (in the continuum) see ~\cite{Pandey2023}, while an approximate construction is given in the supplementary material~\cite{supp}. These states would have  energies $\epsilon_{(y,p)}\approx -\cos p$ and velocities $v_{(y,p)}\approx \sin p$.  It is convenient to use a mapping of the system such that the phase-space domain is $x\in(-2 L_S,0), k \in (0,\pi)$ instead of $x\in(-L_S,0), k \in (-\pi,\pi)$ (see End Matter).  We then form a many-particle state by occupying the $L_S$  levels, $(y,p)$, $y\in(-2L_S,0),~p\in(0,\pi)$, with probabilities   $f_\alpha=f(\epsilon_\alpha)$. The correlation matrix for the so-prepared state will be close to the thermal state in Eq.~\eqref{eq:FT_ini_state}. Let $\phi_\alpha(i,t)=\langle i| \phi(\alpha,t) \rangle$, $i=-L_S+1,-L_S+2,\ldots,L_R$, be the time-evolved state with the full Hamiltonian. Then the correlation matrix at time $t$ is given by $C_{ij}(t)=\sum_{\alpha=1}^{L_S} \phi^*_\alpha(i,t) \phi_\alpha(j,t) f_\alpha$. As explained  in the End Matter,  the eigenvalues of $C^{S,R}(t)$, in either the system or reservoir parts,  can equivalently be obtained respectively from the overlap matrices~\cite{Klich_2006,CMV2011}, defined as
\begin{subequations}
\begin{align}
A^{S}_{\alpha,\nu}(t)=\sum_{j \in S} \sqrt{f_\alpha} \phi^*_\alpha(j,t) \sqrt{f_\nu} \phi_\nu(j,t),  \\
A^{R}_{\alpha,\nu}(t)=\sum_{j \in R} \sqrt{f_\alpha} \phi^*_\alpha(j,t) \sqrt{f_\nu}\phi_\nu(j,t).
\end{align} 
\end{subequations}
In the semiclassical approximation, the wave-packet basis state, $\alpha=(y,p)$, travels to the defect at $x=0$ with speed $v_p$ and is split into two localized packets, a transmitted one with weight $T_p$ and a reflected one with weight $R_p$.  Subsequently, the reflected packet bounces  from the wall and returns to the origin where again it splits into two parts. This keeps repeating and leads eventually to a train of transmitted packets in the reservoir and a  single packet inside the system [see Fig.~\ref{fig:wavepacket}]. This implies that $|\phi_{(y,p)}(t)|^2$ will have support in $s_{\rm max}+1$ cells in the reservoir, with $s_{\rm max}= \lfloor v_p t/(2 L_S)\rfloor$. The weight in each cell is given by $T_pR_p^s$, $s=0,1,\ldots,s_{\rm max}$.  
At time $t=0$, the system overlap matrix, $A^S$ is diagonal while the reservoir one is identically zero. Making the assumption that the wave-packets do not spread significantly, this  implies that the overlap matrices remain diagonal and the diagonal elements can then be  written from the structure of individual wavepackets, discussed above. This leads to the following general form for the eigenvalues of the system and reservoir correlation matrix, respectively, at time $t$~(see \cite{supp} for details), 
\begin{subequations}
\begin{align}
&\xi^{\rm S}(y,p,t)= \nonumber \sum_{x = -2L_S}^{0}   f_p~|\phi_{(y,p)}(x,t)|^2\\ 
&=\nonumber \sum_{s=0}^{\infty}f_pR_{p}^{s} [\theta(-y'-2 s L_S+t v_p) \\
&- \theta(-y'-2 s L_S-2L_S+t v_p)],~~~(-2L_S <y'<0), \label{eq:xi_x<0}\\
&\xi^{\rm R}(y,p,t)= \nonumber \sum_{x = 0}^{\infty} f_p ~|\phi_{(y,p)}(x,t)|^2 \\
&=\nonumber \sum_{s=0}^{\infty}f_p(1-R_{p}^{s+1}) [\theta(-y'-2 s L_S+t v_p) \\
& - \theta(-y'-2 s L_S-2L_S+t v_p)],~~~0<y'<2L_S. \label{eq:xi_x>0}
\end{align}
\end{subequations}
{where $y'$ denotes the final position of the wavepacket (labelled $(y,p$), that maps to the initial position $y$ via ballistic propagation and appropriate boundary conditions imposed by the defect and finite wall.}

In the above, the Heaviside-$\theta$ functions ensure the contribution of all the spatial cells within a distance $v_p t$ where the wavepacket has support. In the reservoir formula, the changed range of $y$ follows after a change of variables. We now show that the above formulas can be related to the expression for phase space density obtained from hydrodynamics.

\textit{Hydrodynamic description of eigenvalues of the Correlation matrix ($\zeta^{S/R}$)--} Evolution of free fermionic system under Eq.~\eqref{set-up} can also be described using Generalized Hydrodynamics~\cite{SKD2024}. In this description, the fermions are treated as a gas of quasiparticles, each specified by their  position and momentum. The equation for the evolution of the quasiparticle phase space density, $n(x,k,t)$, is given by,
\begin{equation}
    (\partial_t + v_k\, \partial_x) n(x,k,t) = 0,
    \label{eq:euler}
\end{equation}
where $v_k = \sin(k)$. The general solution of Eq.~\eqref{eq:euler} on an infinite line is  $n(x,k,t) = n_0(x - v_k t,k)$, where $n_0(x,k)$ is the initial condition. For a mixed state finite temperature initial condition, $n_0(x,k) = f_k \big{(}\theta(-x)-\theta(-x-L_S)\big{)}$. Following  Ref.~\cite{SKD2024}, in the presence of a defect and a finite wall, taking into account the multiple reflections from the defect and the walls,  the solution of Eq.~\eqref{eq:euler} at any  time $t$ can be explicitly constructed (see \cite{supp}). We then find that,  for the system,
\begin{align}
\label{eq:n<0}
    n(x,k,t) = \xi^S(x,k,t),
 \end{align}   
where $\xi^S$ is exactly given by Eq.~\eqref{eq:xi_x<0}. On the other hand, in the reservoir,
\begin{align}
    n(x>0,k>0,&t) = \sum_{s=0}^{\infty}f_kT_k R_k^s \big{(}\theta(-x-2sL_S+t\sin[k])\notag\\
    & - \theta(-x-2sL_S-2L_S + t\sin[k])\big{)}.
    \label{eq:n_res}
\end{align}
\begin{figure}
    \centering
\includegraphics[width=\linewidth]{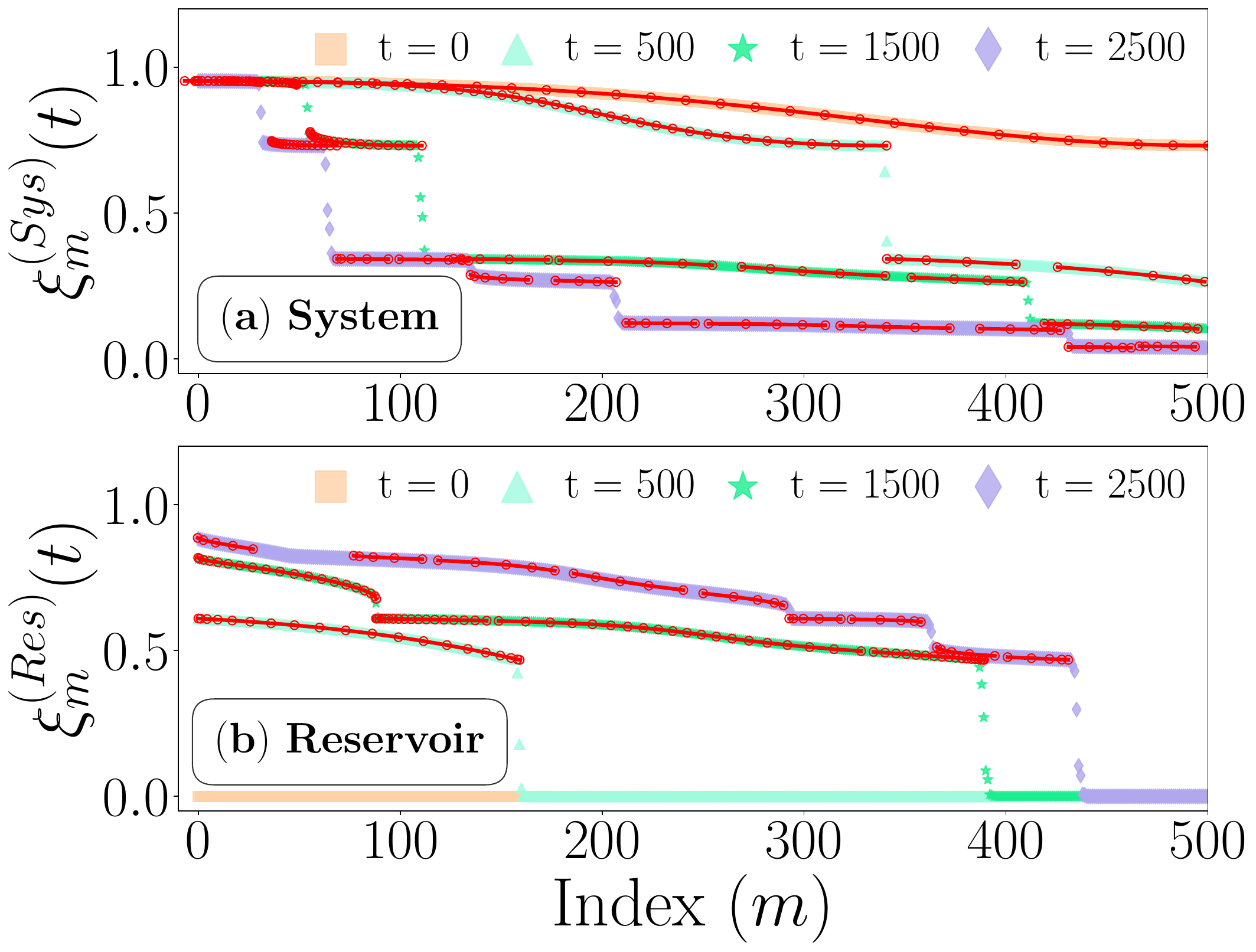}
    \caption{ Eigenvalue profile plotted at different times for (a) the system part of the correlation matrix and (b) the reservoir part of the correlation matrix. The symbols correspond to numerically obtained eigenvalues while the red dotted line are ones predicted by hydrodynamics. 
    The conformal defect with $g_c= 0.4$ was considered with other parameters, $\beta = 1$, $\mu = 2$ and $L_S=500$.}
    \label{fig2}
\end{figure}
It then follows that, for $x>0$,
\begin{align}
&\sum_{m=0}^\infty n(x+2 m L_S,k,t) \nonumber \\
=& \sum_{m=0}^\infty \sum_{s=0}^{\infty}f_kT_k R_k^s \big{(}\theta(-x-2(s+m)L_S+t\sin[k])\notag\\
    & - \theta(-x-2(s+m) L_S-2L_S + t\sin[k])\big{)}.
\end{align}
Making a change of variable $s+m=j, s-m=l$ we get 
\begin{align}
& \sum_{j=0}^\infty \sum_{l=-j/2}^{j/2}f_kT_k R_k^{(j+l)/2} \big{(}\theta(-x-2jL_S+t\sin[k])\notag\\
    & - \theta(-x-2j L_S-2L_S + t\sin[k])\big{)}.
\end{align}
After performing the sum over $j$, allowing only integer values, we find,   
on comparing this with Eq.~\eqref{eq:xi_x>0},
\begin{align}
 \sum_{m=0}^\infty n(x+2 m L_S,k,t)=   \xi^R(x,k,t).
\label{eq:xi_x>0-2}
\end{align}
This corresponds to partitioning the domain $x =(0,\infty)$ into boxes of size $(0,2L_S)$ and adding the phase space densities of all these boxes gives us the reservoir eigenvalues.   Thus, we now have a hydrodynamic interpretation for the eigenvalues of the correlation matrix for both  system and reservoir. 

To further elucidate the connections to GHD, we take note of  the observation in Ref~\cite{Pandey2023} that the diagonal elements of the correlation matrix in the wave-packet basis give exactly the coarse-grained Wigner function, which is in fact the quasiparticle phase space density $n(x,k)$ for the system ~\cite{essler2023}. For the reservoir, the modified phase-space density is equivalent to the diagonal elements of the correlation matrix in the wave-packet basis.

\emph{Comparison of eigenvalues from exact numerics with hydrodynamic predictions---} The eigenvalues  given by Eq.~\eqref{eq:xi_x<0} and \eqref{eq:xi_x>0} are labelled by the indices  $x$ and $k$ whereas the exact eigenvalues obtained from diagonalization of the correlation matrix are an ordered set depending on a single integer index, say $\ell$. It turns out that one can precisely relate these two (see End Matter). Using this correspondence, we  compute  the hydrodynamic predictions for the eigenvalues of the correlation matrix and compare them with the exact numerical ones. The results are plotted in Fig.~\ref{fig2}, which shows the eigenvalue profile, at different times,  for both the system and the reservoir parts of the correlation matrix.  A finite temperature mixed state initial condition was considered  for the case of the conformal defect.  We see excellent agreement between the numerically obtained eigenvalues and those obtained from hydrodynamics.
\begin{figure*}
    \centering
    \includegraphics[width=\linewidth]{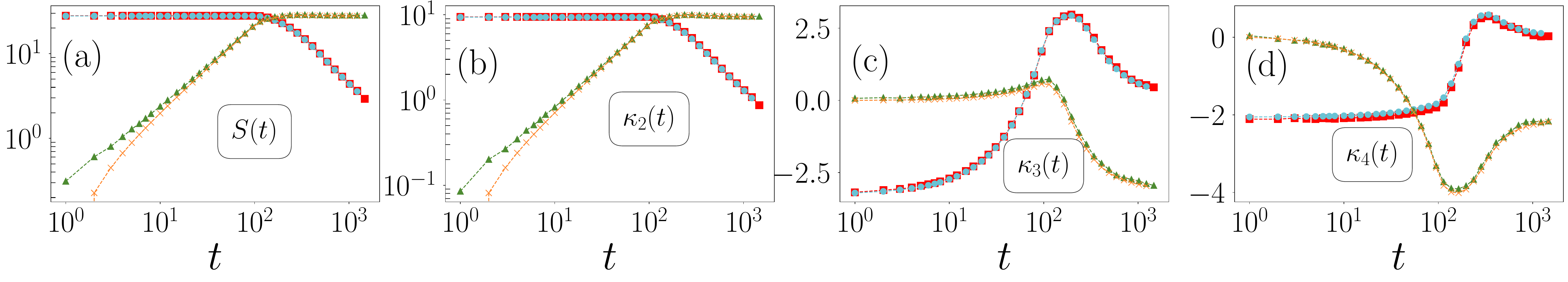}
    \caption{Comparison between vN entropy and cumulants obtained using exact numerics and hydrodynamical equations. The red squares and the blue circles correspond to the results from exact microscopic and hydrodynamic, respectively, for the system part. Green triangles and orange crosses correspond to the results from exact microscopics and hydrodynamics, respectively, for the reservoir part. Here, $L_S = 50$, $\beta = 1$, $\mu = 1$ and $g_c = 0.4$. }
    \label{fig:Numerics}
\end{figure*}

 \textit{Entropy and number fluctuations --} We now use these  analytic expressions for the  eigenvalues of the correlation matrix from hydrodynamics to determine the von Neumann entropy and the full counting statistics.
 In Fig.~\ref{fig:Numerics} we show the comparison between the numerical results for microscopically obtained vN entropy and FCS using Eq.~\eqref{eq:vN_entropy} and Eq.~\eqref{eq: FCS}, respectively, and those obtained from generalized hydrodynamics using Eq.~\eqref{eq:n<0} and Eq.~\eqref{eq:xi_x>0-2}. 
The results are for a finite temperature mixed initial state with $\beta = 1$, $\mu = 1$, for the case of a conformal defect with $g_c/g = 0.8$. 
 We see a remarkable agreement between exact microscopics and  hydrodynamics for both von Neumann entropy and FCS, for both system and reservoir, except at very early times.  As discussed in SM~\cite{supp}, the agreement in fact holds for other point defect setups such as hopping defect, density defect, etc., and for any  values of $\beta$ and $\mu$.  
 
 \textit{Need for a modified YY formula}-- We note that  the naive approach of computing the reservoir vN-entropy using the YY formula fails because this essentially treats the multiple peaks of the wavefunction as incoherent parts. However, a key point is that all of these wavepackets in the reservoir are produced from the same wavepacket in the system, so they are all coherent with one another and should be considered all together as one single wavepacket that has these multiple peaks as an ``internal'' structure.  Since this ``internal'' structure of this wavepacket is coherent, it does not contribute to the vN entropy of the reservoir.

 \begin{figure}
    \centering
    \includegraphics[width=0.9\linewidth]{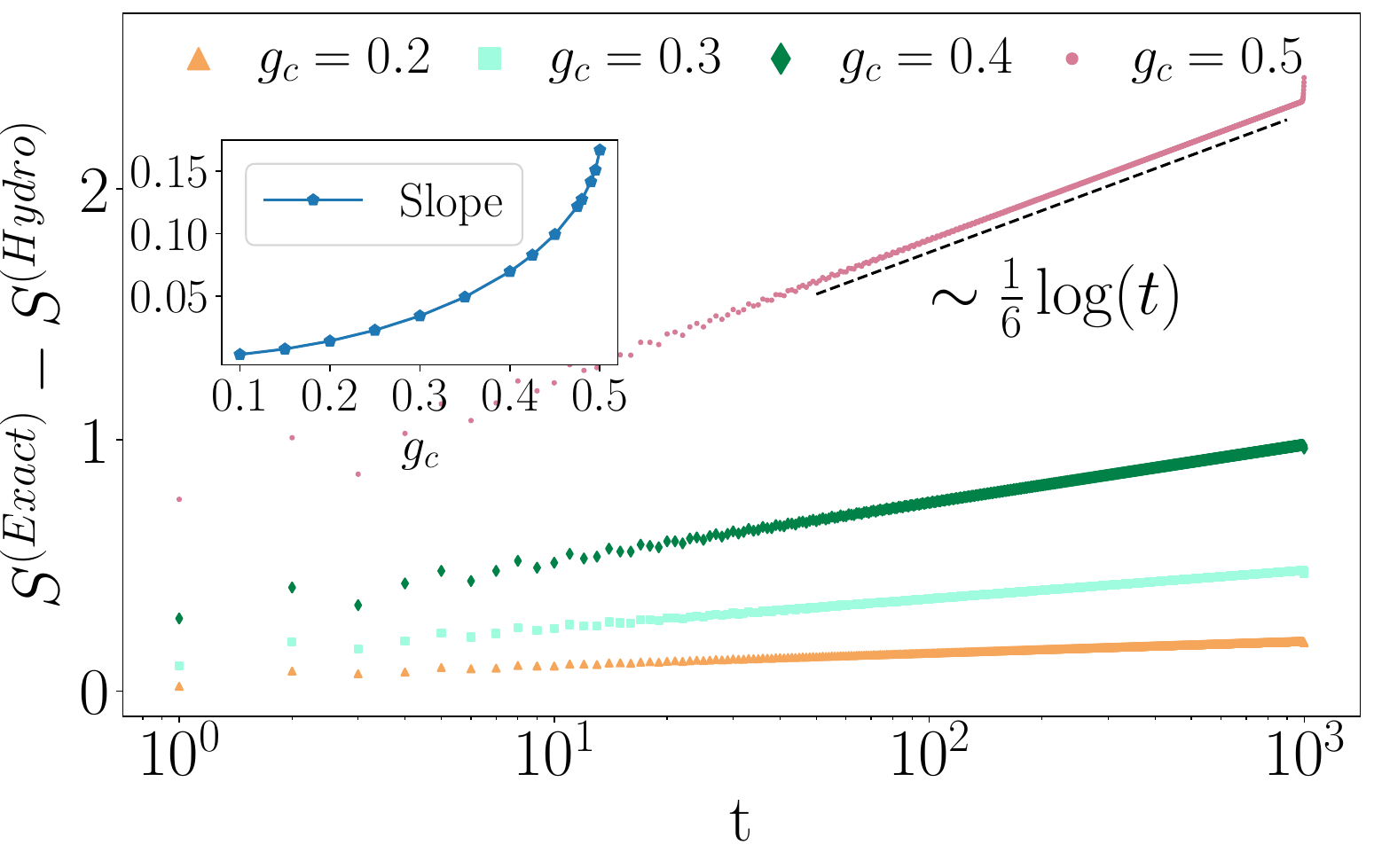}
    \caption{Difference between the exact vN entropy (numerically computed) and the hydrodynamic entropy as a function of time for a setup with  a conformal defect with different coupling strengths at $\beta = \infty$ and $L_S = 500$. A logarithmic growth is seen. For $g_c=1/2=g$ , the difference   $~\frac{1}{6}\log(t)$~\cite{logt_growth_analytical}, which is the well known result for entanglement growth in the defectless case. Inset: The prefactor of the $\log(t)$ growth is plotted as a function of defect strength $g_c$. }
    \label{fig:log_corr}
\end{figure}

 We expect that hydrodynamics provides the leading order semiclassical approximation for the entropy. However, there are interesting quantum fluctuations that are not accounted for, and this in fact shows up nicely in the eigenvalue spectrum. In Fig.~\ref{fig2}, we see a small fraction of eigenvalues that are transitioning from the $R^{s}$ level to the $R^{s+1}$ level, that are not predicted by the hydrodynamic formula. These  eigenvalues are responsible for logarithmic corrections to the vN entropy, as well as particle number fluctuations.
This is demonstrated in Fig.~\ref{fig:log_corr} where we plot the difference of the 
exact and hydrodynamic entropies for the case of the conformal defect with different values of $g_c$ and with $\beta=\infty$.   

\textit{Summary and outlook --} To summarize, we have considered an expanding free fermionic gas in the presence of a point defect with a finite temperature mixed initial state. For such a setup, using the eigenvalues of the correlation matrix, we can compute quantities such as vN entropy and FCS. In this letter, we provide a heuristic hydrodynamic theory, based on the semiclassical wavepacket picture,  to obtain analytic expressions for the correlation matrix eigenvalues.  We find excellent agreement with eigenvalues obtained from exact numerics, except for a small subset.  Subsequently, the vN entropy and the FCS of both the system and reservoir are also obtained from the hydrodynamic approach. Our results thus provide a justification for the Yang-Yang formula and show the  modifications required for the reservoir. Finally, the results on FCS provide an extension of the Levitov-Lesovik results to the case of a finite system coupled to a bath and in the  transient regime. 

Our findings help us to resolve the mismatch of the reservoir hydrodynamic entropy with the exact numerics, discussed in \cite{SKD2024}. Thus, such a resolution addresses the issue of the non-monotonic growth of the reservoir's entanglement entropy, which has been widely discussed in the context of black holes and Hawking's semiclassical calculations~\cite{hawking,aa_review,Mathur_2009}.

It is worth noting that such setups are suitable to study dynamics of various observables in experimental platforms such as cold-atoms in optical lattices, solid state magnets, qubit devices etc. \cite{GHD_experiment_review,ultra_cold_atoms_review,Cheneau2012,Jurcevic2014,Lanyon2017}, thus making our findings experimentally relevant. It will be interesting to study the hydrodynamics description of the correlation matrix eigenvalues in the presence of two or multiple baths. Here, we have only considered a nearest-neighbor hopping. It is interesting and challenging to extend this idea to long-range hopping models~\cite{RMP_LR,knap2020,dhawan2024}. \\

\section*{Acknowledgements}
A.D. acknowledges support from the J.C.~Bose Fellowship (JCB/2022/000014) of the Science and Engineering Research Board, Department of Science and Technology, Government of India. D.A.H. was supported in part by NSF QLCI grant OMA-2120757.  We also acknowledge support from the Department of Atomic Energy, Government of India, under Project No. RTI4001. 

\bibliography{references}

\clearpage


\setcounter{equation}{0}
\setcounter{figure}{0}
\renewcommand{\theequation}{EM\arabic{equation}}
\renewcommand{\thefigure}{EM\arabic{figure}}
\onecolumngrid

\section*{\underline{End Matter}}

\section{Mapping system domain from $(-L_S,0),(-\pi,\pi)$ to $(-2L_S,0),(0,\pi)$ }

Here we discuss the equivalence between formulating the system on the interval $(-L_S,0)$ and on the unwrapped interval $(-2L_S,0)$. In the original description, the momentum quantum number spans the full Brillouin zone, $k\in[-\pi,\pi]$, so that positive and negative momenta correspond to right- and left-moving excitations, respectively. The negative-$k$ sector therefore represents modes propagating toward decreasing spatial coordinates within the interval $(-L_S,0)$. An equivalent description is obtained by extending the system to $(-2L_S,0)$ and restricting the momentum label to $k\in[0,\pi]$. In this unwrapped representation, the degrees of freedom that were previously associated with negative momenta are reinterpreted as positive-momentum particles originating from the additional spatial segment $(-2L_S,-L_S)$. Thus, a left-moving excitation in the original formulation is mapped to a right-moving excitation on the enlarged interval, with the direction of propagation encoded in the spatial origin of the mode rather than in the sign of its momentum. The correspondence is one-to-one, ensuring that no degrees of freedom are lost or introduced in passing from one representation to the other. In the new representation, on reflection from the defect at the origin, a particle starts again from $x=-2L_S$ with the same momentum.

\section{Eigenvalues from overlap matrix, relation between system and reservoir eigenvalues}

Consider the matrices, $M^D$ ($D=S/R$), with matrix elements $M^{D}_{i,\alpha}=\sqrt{f_\alpha} \phi_\alpha(i)$, with $\alpha=1,2,\ldots,L_S$, while $i=-L_S+1,-L_S+2,\ldots,0$  for $D=S$ and $i=1,2,\ldots,L_R$  for reservoir. We assume always that $L_R>L_S$. Using Schmidt decomposition, it follows that
\begin{align}
M^D=U^D \Lambda^D [V^D]^\dagger,   
\end{align}
where $V^D$ is a $L_S \times L_S$ unitary matrix, $U^S, U^R$ are respectively $L_S \times L_S$ and $L_R \times L_R$ unitary  matrices. Also, $\Lambda^S$ is a $L_S \times L_S$ diagonal matrix with entries $\sigma^S_s$ ($s=1,2,\ldots,L_S$), while $\Lambda^R$ is a $L_R \times L_S$ matrix with only the top $L_S \times L_S$ block non-zero and diagonal with non-zero elements $\sigma^R_s$  ($s=1,2,\ldots,L_S$). 

Then it is clear  that $C^D=M^D [M^D]^\dagger$ and $A^D=[M^D]^\dagger M^D$ will have the same set of non-zero eigenvalues, $\xi^D_s=(\sigma^D_s)^2$.  This completes the proof. 

Note that $A^S+A^R=M^S [M^S]^\dagger + M^R [M^R]^\dagger$. Taking matrix elements, one can see that 
$A^S_{\alpha, \nu}+A^R_{\alpha,\nu}= f_\alpha \delta_{\alpha, \nu}$.

\section{Eigenvalue correspondence derivation}

We now describe the correspondence  between the exact eigenvalues with those given from hydrodynamics. First consider the reservoir eigenvalues which take the values  $f_k(1 - R_k^{s+1})$.  From Eq.~\eqref{eq:xi_x>0}, we see that for any fixed $s$, the range of allowed $k$ modes is determined by the conditions $ v_k t-2(s+1)L_S < y  < v_k t -2 s L_S$ and $0<y<2 L_S$. Two possibilities arise: 
\begin{enumerate}
\item $v_k t -2 s L_S < 2L_s$ and the the allowed range of $k$ is
\begin{align}
k\in[\sin^{-1}(\frac{2sL_S}{t}),\sin^{-1}(\frac{2(s+1)L_S}{t})]
    \bigcup [\pi - \sin^{-1}(\frac{2(s+1)L_S}{t}),\pi - \sin^{-1}(\frac{2sL_S}{t})] \notag
    \end{align}
 For any $k$ in the above ranges, all values of $y$ in the range $ (0,v_k t -2 s L_S)$ lead to the same value of $\xi^R$. There are thus $v_k t -2 s L_S$ possible $y$. This leads to a relation between the eigenvalue index $\ell$ and the momentum index $k$, namely we have:
 \begin{align}
 \label{eigenval_corres1}
   \xi_\ell= (1-R_k^{s+1}),~ {\rm with}~~d \ell =[v_k t -2 s L_S] \frac{dk}{\pi}.   
 \end{align}
 
 \item $0 < v_k t-2(s+1)L_S$, the allowed range of $k$ is:  
\begin{align}
&k\in[\sin^{-1}(\frac{2(s+1)L_S}{t}),\sin^{-1}(\frac{2(s+2)L_S}{t})]
    \bigcup [\pi - \sin^{-1}(\frac{2(s+2)L_S}{t}),\pi - \sin^{-1}(\frac{2(s+1)L_S}{t})]
    \label{eq:eig1}
\end{align}
 In this case all $y$ in the range $(v_k t-2(s+1)L_S, 2L_S)$ give the same $\xi^R$ and this  leads to the relation 
 \begin{align}
 \label{eigenval_corres2}
   \xi_\ell= (1-R_k^{s+1}),~ {\rm with} ~~d \ell=[2(s+2) L_s -v_k t] \frac{dk}{\pi}.   
 \end{align}
 \end{enumerate}

\section{Computation of VN entropy and FCS using the hydrodynamic description of eigenvalues of the correlation matrix}

Using the relation between hydrodynamic phase space density and correlation matrix eigenvalues in Eq.~\eqref{eq:n<0}, \eqref{eq:xi_x>0-2}, we can compute the vN entropy and FCS. Thus, the formulas to calculate vN entropy from the system and reservoir part using hydrodynamics is the following:

\begin{eqnarray}
S^{\rm{S}}_{(\rm{Hydro})} &=&- \int_{-2L_S}^0 dx \int_{0}^{\pi} \frac{dk}{2\pi} \bigg[ \xi^S(x,k,t)\log \xi^S(x,k,t) + 
\left(1 - \xi^S(x,k,t)\right)
\log\left(1 - \xi^S(x,k,t)\right) \bigg]\\
S^{\rm{R}}_{(\rm{Hydro})} &=& -\int_0^{2L_S} dx \int_{0}^{\pi} \frac{dk}{2\pi} \bigg[ \xi^R(x,k,t)\log \xi^R(x,k,t) + 
\left(1 - \xi^R(x,k,t)\right)
\log\left(1 - \xi^R(x,k,t)\right) \bigg].
\end{eqnarray}
Similarly, we give the hydrodynamic description for computing FCS using Eq.~\eqref{eq:n<0}, \eqref{eq:xi_x>0-2} which takes the following form:
\begin{eqnarray}
\mu^{\rm{S}}_{(\rm{Hydro})} (\lambda) &=& \int_{-2L_S}^0 dx \int_{0}^{\pi} \frac{dk}{2\pi} \bigg[ 
\log\left(1 +(e^{\lambda}-1) \xi^S(x,k,t)\right) \bigg]\\
\mu^{\rm{R}}_{(\rm{Hydro})} (\lambda) &=& \int_{0}^{2 L_S} dx \int_{0}^{\pi} \frac{dk}{2\pi} \bigg[ 
\log\left(1 +(e^{\lambda}-1) \xi^R(x,k,t)\right) \bigg].
\end{eqnarray}

\clearpage

\setcounter{section}{0}
\setcounter{figure}{0}
\setcounter{table}{0}
\setcounter{equation}{0}
\renewcommand{\thesection}{S\arabic{section}}
\renewcommand{\thefigure}{S\arabic{figure}}
\renewcommand{\thetable}{S\arabic{table}}
\renewcommand{\theequation}{S\arabic{equation}}

\begin{center}
  \textbf{\large Supplementary Material:\\ Hydrodynamic description of correlation matrix eigenvalues in a finite temperature expanding free fermionic gas  }
\end{center}

\vspace{1em}
\twocolumngrid

\section{Details of mixed state calculation}
\label{sec:mixed}
In this section, we will discuss the details of calculating the mixed state finite temperature von Neumann entropy using (i) exact numeric and (ii) generalised hydrodynamics. The system that we work on has the Hamiltonian of the form,
\begin{equation}
    \hat{H}= \hat{H}_S + \hat{H}_R + \hat{V}
    \label{supp_eq:Ham}
\end{equation}
with,
\begin{align}
&\hat{H}_S = -g\sum_{i = -L_S+1}^{-1} [\hat{c}^\dagger_{i}\hat{c}_{i+1} + \hat{c}^\dagger_{i+1}\hat{c}_i] \label{supp_eq:HS}\\
    &\hat{H}_R = -g\sum_{i = 1}^{L_R-1} [\hat{c}^\dagger_{i}\hat{c}_{i+1} + \hat{c}^\dagger_{i+1}\hat{c}_i]\\
    &\hat{V} = -g_c[\hat{c}^\dagger_{0}\hat{c}_{1} + \hat{c}^\dagger_{1}\hat{c}_0] +\sqrt{g^2-g_c^2}[\hat{c}^\dagger_{0}\hat{c}_{0} + \hat{c}^\dagger_{1}\hat{c}_1]
\end{align}
where $\hat{H}_S$ and $\hat{H}_R$ are the system and the reservoir Hamiltonian respectively and $\hat{V}$ represents the coupling between the system and the reservoir. We consider the case of a conformal defect with $g_c<g$ in which case the reflection and the transmission probabilities of the underlying scattering problem are independent of the momentum modes~\cite{SKD2024}.
\subsection{Exact Numerics}
\label{subsec:en_mixed}
To compute the von Neumann entropy for a finite temperature mixed state, we will first evolve the correlation matrix,
\begin{equation}
    C^{}_{ij}(t) = \langle\hat{c}_i^\dagger(t)\hat{c}_j(t)\rangle
\end{equation}
using the von Neumann evolution equation,
\begin{equation}
    C(t) = e^{iht} C(t = 0) e^{-iht},
    \label{supp_eq:vN_evol1}
\end{equation}
where $h$ is the single particle Hamiltonian defined as,
\begin{equation}
    \hat{H}=\sum_{i,j=-L_S+1}^{L_R} h_{ij} \hat{c}^{\dagger}_i \hat{c}_j,
\end{equation}
and $C(t = 0)$ is the finite temperature correlation matrix at $t = 0$. To construct the initial mixed state correlation matrix, $C_{ij}(t = 0) = \langle\hat{c}^\dagger_i(0)\hat{c}_j(0)\rangle$, consider the system Hamiltonian , $\hat{H}_S$ given by Eq.~\eqref{supp_eq:HS}. Diagonalising $\hat{H}_S$ we get,
\begin{equation}
    \hat{H}_S = \sum_{q = 1}^{L_S} \lambda^S_q\hat{d}^\dagger_q\hat{d}_q,
\end{equation}
 where $\{\lambda^S_q\}$ are the eigenvalues of $\hat{H}_S$ and $\hat{c}_i = \sum_{q = 1}^{L_S}U_{iq}\hat{d}_q$. The unitary matrix $U_{ij}$ is composed of the columns that are eigenvectors of $\hat{H}_S$. Writing $C(0)$ in the diagonal basis of $\hat{H}_S$ we get,
\begin{align}
    C_{ij}(0) &= \sum_{q,p = 1}^{L_S}U^*_{qi}U_{jp}\langle\hat{d}^\dagger_q \hat{d}_p\rangle\\
    & = \sum_{q = 1}^{L_S}U^*_{qi}U_{jq}f(\lambda_q)
\end{align}
where we have defined $\langle\hat{d}^\dagger_q \hat{d}_p\rangle\equiv f(\lambda_q)\delta_{q,p}$ with the Fermi distribution function,
\begin{equation}
    f(\lambda_q) = [e^{\beta(\lambda_q - \mu)}+1]^{-1}.
\end{equation}
Once $C_{ij}(0)$ is constructed, it can be evolved using Eq.~\eqref{supp_eq:vN_evol1}. To compute the von Neumann entropy, we diagonalise the restricted correlation matrix, which is given by $[C_{ij}]_{i,j = -L_S+1}^{0}(t)$ for the system part and $[C_{ij}]_{i,j = 1}^{L_R}(t)$ for the reservoir part. This gives us the eigenvalues $\{\zeta_{i}^D\}$ with $D=S$ for the system and $D=R$ for the reservoir. Using these eigenvalues, the finite-temperature mixed-state von Neumann entropy can be computed as,
\begin{equation}
\label{supp:SvN}
    S_{v}^D(t) = -\sum_{i = 1}^{L_S}\Big{[}\zeta^D_i\log[\zeta^D_i] + [1-\zeta^D_i]\log[1-\zeta^D_i]\Big{]}.
\end{equation}
Note that for the reservoir part, only $L_S$ number of eigenvalues will be non-zero. Hence, even for $D=R$, the summation runs over only $L_S$ eigenvalues.

\section{Eigenvalues of the system and reservoir correlation matrix using a heuristic wave-packet description}

\subsection{Construction of approximate orthonormal wavepacket basis states}

In this subsection, we will discuss in detail the construction of the Gaussian wavepacket basis states. We consider a system of size $L_S$ with $N$ particles and an empty reservoir of size $L_R\gg L_S$. The $N$ particles are in a Slater determinant state where the single-particle states are localised wave packets $\phi_\alpha(x,t)$, with $\alpha = 1,2,..., N$, where $\alpha \equiv (y,p)$ labels the mean position $y$ and mean momentum $p$. We define the initial state of the particles using Gaussian wave packets as,
\begin{equation}
    \phi_{(y,p)}(x,t=0) = \mathcal{N}e^{-ipy}e^{-\frac{(x-y)^2}{2(\Delta x)^2}}
\end{equation}
where $\mathcal{N}$ is the normalization factor, $y$ is the initial position of the wave packet, $p$ is the group velocity, and $\Delta x$ is the standard deviation of the Gaussian wave packet that controls the initial spread.

To construct an approximate orthonormal basis with the localised wavepacket, $\phi_{(y,p)}(x,t)$, that is localised around the mean position $y$, within a width of $\Delta x$, and around mean momentum $p$ within a width of $\Delta k$, we require these states to satisfy the uncertainty relation, $\Delta x \Delta k \gtrsim 2\pi$. To accomplish this, we define $\Delta x = \frac{L_S}{K}$ and $\Delta k = \frac{2\pi K}{L_S}$ where $K$ is an integer which denotes the number of partitions between $x = -L_S$ to $x = 0$. Given these $\Delta x$ and $\Delta k$, we take $y = n\Delta x$ and $p = m\Delta k$ where $n = 0,1,\cdots,K-1$, $m = -\frac{L_S}{2K}+1, -\frac{L_S}{2K}+2 ,\cdots,\frac{L_S}{2K}$. To define the initial condition with $N$ particles, we choose $ N$ pairs from the set $(\{y\},\{p\})$. 

\subsection{Correlation matrix eigenvalues from wavepacket description}

Here we derive Eqs.~(8a,8b) of the main text.
As discussed in the main text, at $t = 0$ the system overlap matrix $A^S$ is diagonal and the reservoir overlap matrix, $A^R$, is $0$. Assuming that the wavepackets do not spread significantly, the overlap matrix is approximately diagonal, and the diagonal elements, which are also the eigenvalues, are given by (for large $L_S$ we ignore the difference with $L_S-1$),
\begin{equation}
    \xi^S(y,p,t) = f_p\sum_{x = -L_S}^0 |\phi_{(y,p)}(x,t)|^2
\label{eq_supp:xiS}
\end{equation}
for the system part, and 
\begin{equation}
\label{eq:psiR}
    \xi^R(y,p,t) = f_p\sum_{x = 0}^\infty |\phi_{(y,p)}(x,t)|^2
\end{equation}
for the reservoir part with  $-L_S<y\le 0$ and $p\in[-\pi,\pi]$ and 
$f_p$ is the Fermi distribution function with energy $\epsilon_{(y,p)} \approx \cos(p)$. At $t = 0$, it is easy to see that $\xi^S(y,p) = f_p$ and $\xi^R(y,p) = 0$. 

We will now use the semiclassical picture to compute the time-evolved wave packets, $\phi_{(y,p)}(x,t)$.  A wavepacket initialized in the system at the position $y$ with momentum $p$ starts moving right towards the defect at $t = 0$. Upon reaching the defect, it is scattered, and a part of the wave packet is transmitted to the reservoir, and the remaining part is reflected back into the system. The transmitted wave packet moves deeper into the reservoir, whereas the reflected wave packet moves towards the finite wall, gets reflected, and approaches the defect again. Upon hitting the defect, a part of the wave packet is again transmitted to the bath, and the rest is reflected back to the system. This process keeps repeating. At every scattering event, the transmitted wavepacket's normalzation  is reduced by a factor $T_p$ compared to the incident wave-packet, while the reflected one has a normalization reduced by a factor of $R_p$. 

We now use the representation defined in End Matter (Sec. I) where we consider only right movers and the system phase space is specified as $p \in (0,\pi), y \in (-2 L_S,0)$. The semiclassical picture presented  for the wavepacket evolution in the previous paragraph means that we can write any evolved  wavepacket in the form:
\begin{align}
&|\phi_{(y,p)}(x,t)|^2 = \sum_{s=0}^\infty |\phi_{(y,p)}(x-v_p t+2sL_S,0)|^2  R_p^s \nonumber\\
&\big{[}\theta(-x-2 s L_S+t v_p) - \theta(-x-2 s L_S-2L_S+t v_p)\big{]}, \notag \\
~~~&{\rm for}~ -2 L_S <x <0.  \label{eq:sys}
\end{align}
Only one term in the sum will contribute. The $\theta$ functions arise from the observation that, at time $t$, the wavefunction is peaked at $x=y+v_pt-2s L_S$. The condition that $y$ is in the range $-2 L_S < y <0$ can be incorporated through the factor $\theta(-y)-\theta(-y-2L_S)$ and replacing $y$ by $x-v_p t+2 s L_S$ gives us the  form in Eq.~\eqref{eq:sys}.

We next need to sum Eq.~\eqref{eq:sys} over $x$ [recall Eq.~\eqref{eq_supp:xiS} along with unwrapping discussed in Sec.~I discussed in End Matter].  Summing over $x$ in the system gives us
\begin{align}
&\xi^S(y,p,t)=f_p\sum_{x=-2 L_S}^0 |\phi_{(y,p)}(x,t)|^2 \notag \\ &= \sum_{s=0}^\infty   f_p R_p^s \sum_{x=-2 L_S}^0  |\phi_{(y,p)}(x-v_p t+2sL_S,0)|^2   \nonumber\\
&\big{[}\theta(-x-2 s L_S+t v_p) - \theta(-x-2 s L_S-2L_S+t v_p)\big{]}, \notag \\
~~~&{\rm for}~ -2 L_S <y <0.  \label{eq:sys2-v0}
\end{align}
For a specified $p$ and a time $t$, only one of the terms (say $s'$) in the summation over $s$ is non-vanishing. This corresponds to the single localized surviving packet after $s'$ reflections at the defect (see Fig.~(1) of main text). Let $y'$ denote the peak position of this leftover bump. Noting that $\phi_{(y,p)}(x,0)$ is normalized, we then get,   
\begin{align}
&\xi^S(y,p,t)=  \sum_{s=0}^\infty   f_p R_p^s \big{[}\theta(-y'-2 s L_S+t v_p) \notag \\ &~~~~~~- \theta(-y'-2 s L_S-2L_S+t v_p)\big{]}, \notag \\
~~~&{\rm for}~ -2 L_S <y' <0,  \label{eq:sys2}
\end{align}
where again the $\theta$ functions ensure that for given $t,y,p$, only one $s$ is selected, corresponding to the right number of reflections that lead to the particle being in the box $(-2L_S,0)$. This recovers the result in Eq.~(8a) in the main text.

Now, let us consider the reservoir part. Since the evolution is unitary, $\sum_{x = -2L_S}^{\infty}|\phi_{(y,p)(x,t)}|^2 = 1$ for all times and states defined by $(y,p)$. Thus,
\begin{equation}
\label{eq:prob1}
    \sum_{x = 0}^{\infty}|\phi_{(y,p)}(x,t)|^2 = 1 - \sum_{x = -2L_S}^{0}|\phi_{(y,p)}(x,t)|^2.
\end{equation}
Using  Eq.~\eqref{eq:prob1} in Eq.~\eqref{eq:psiR} we get,
\begin{align}
\label{eq:xiR_steps}
    \xi^R(y,p,t) 
    & = f_p\Big{[}1 - \sum_{x = -2L_S}^{0}|\phi_{(y,p)}(x,t)|^2\Big{]} \nonumber \\
    & = f_p\sum_{s = 0}^{\infty} (1 - R_p^s)[\theta(-y'-2 s L_S+t v_p)\nonumber\\
&- \theta(-y'-2 s L_S-2L_S+t v_p)],\nonumber \\&~~~~~(-2L_S <y'<0)
\end{align}
where we used Eq.~\eqref{eq:sys2} and the fact that 
$\sum_{s=0}^\infty [\theta(-y'-2 s L_S+t v_p)- \theta(-y'-2 s L_S-2L_S+t v_p)] =1$. Next, we substitute $y' \to y' - 2L_S$ and $s \to s+ 1$ in Eq.~\eqref{eq:xiR_steps} to get
\begin{align}
    &\xi^R(y,p,t) = f_p\sum_{s = 0}^{\infty} (1 - R_p^{s+1})[\theta(-y'-2 s L_S+t v_p)\nonumber\\
&- \theta(-y'-2 s L_S-2L_S+t v_p)],~~~(0 <y'<2L_S),
\end{align}
which is Eq.~(8b) in the main text.

\section{Phase space densities from generalized hydrodynamics}
\label{subsec:ghd_mixed}

As discussed in the main text, for the free fermionic Hamiltonian given by Eq.~\eqref{supp_eq:Ham},  generalized hydrodynamics (or simply the collisionless Boltzmann equation) gives the evolution equation of the phase space density, $n(x,k,t)$, of the corresponding quasiparticles as,
\begin{equation}
    (\partial_t + \sin[k]\partial_x)n(x,k,t) = 0.
    \label{supp_eq:Euler}
\end{equation}
This equation needs to be solved after incorporating the fact that quasiparticles with momentum $k$ are reflected or transmitted at the defect with probabilities $R_k$ and $T_k$, respectively. In addition,  they are reflected at the wall $x=-L_S$. 
It is easy to extend  the approach followed in Ref.~\cite{SKD2024}  for the  zero-temperature setup with an initial domain wall configuration. In Ref.~\cite{SKD2024}, the solution of Eq.~\eqref{supp_eq:Euler}, taking into account multiple refelections from the defect and fixed wall, was  stated as,
\begin{align}
\label{supp_eq:Hrdro_sol}
    n(x,k,t) &= \Big{[}n(x<0,k<0,t) + n(x<0,k>0,t)\Big{]}\theta(-x) \notag\\
    &+ \Big{[}n(x>0,k>0,t)\Big{]}\,\theta(x),
\end{align}
where,
\begin{align}
    \label{supp_eq:n_sys}
    n(x<0,k>0,t) =&\sum_{s=0}^{\infty}R_k^s \big{(}n_0(x+2 s L_S-t\sin[k]) \notag\\
    & + n_0(-x-2s L_S-2 L_S + t\sin[k])\big{)},\\
  \label{supp_eq:n_sysm}  n(x<0,k<0,t) =& \sum_{s=0}^{\infty}R_k^{s+1} n_0(-x+2s L_S+t\sin[k]) \notag\\
    & + R_k^s n_0(x-2s L_S - t\sin[k]),\\
    n(x>0,k>0,t) =& \sum_{s=0}^{\infty}T_k R_k^s \big{(}n_0(x+2s L_S-t\sin[k])\notag\\
    & + n_0(-x-2s L_S-2 L_S+t\sin[k])\big{)},
    \label{supp_eq:n_Res}
\end{align}
and the initial phase space density was taken to be
\begin{equation}
    n_0(x,k) = \theta(-x) - \theta(-x-L_S) \hspace{1em}\textrm{for}\,\, k\in(-\pi,\pi). \label{eq:n0}
\end{equation}
This initial condition  corresponds to the domain wall initial condition. Note that $s$ denotes the number of reflections off the finite wall. 

For the case where the system is  initially in thermal equilibrium at inverse temperature $\beta$ and chemical potential $\mu$, a quasiparticle occupies an energy level $\epsilon_k$ with probablity $f_k$, the Fermi distribution factor. Hence 
Eq.~\eqref{eq:n0} is replaced by, 
\begin{equation}
    n_0(x,k) = f_k [\theta(-x) - \theta(-x-L_S)] \hspace{1em}\textrm{for}\,\, k\in(-\pi,\pi). \label{eq:n0T}
\end{equation}
This, used in Eqs.~(\ref{supp_eq:n_sys}-\ref{supp_eq:n_Res}),  gives the solution for the evolution of the phase space density $n(x,k,t)$ with thermal initial conditions.
We now show that this leads to the forms, Eq.~(10) and Eq.~(11) in the main text. Using Eq.~\eqref{eq:n0T} in Eqs.~(\ref{supp_eq:n_sys}-\ref{supp_eq:n_Res}), we can write the phase space densities in the following forms,
\begin{align}
    \label{supp_eq:n_sys_theta}
    n(x<0,k>0,t) =&\sum_{s=0}^{\infty}f_k\,R_k^s \Big{[}\theta(-x-2 s L_S+t\sin[k]) \notag\\
    &- \theta(-x-2 s L_S-L_S+t\sin[k]) \notag \\
    & + \theta(x+2s L_S+2 L_S -t\sin[k]) \notag \\
    &-\theta(x+2s L_S+L_s-t\sin[k])
    \Big{]},\\
  \label{supp_eq:n_sysm_theta}  n(x<0,k<0,t) =& \sum_{s=0}^{\infty}f_k \,R_k^{s+1} \Big{[}\theta(x-2s L_S-t\sin[k]) \notag\\
  &-\theta(x-2s L_S-L_S-t \sin[k]) \Big{]}\notag \\
    & + f_k R_k^s \Big{[}\theta(-x+2s L_S + t\sin[k])\notag \\
    &-\theta(-x+2s L_S-L_S+t \sin[k]) \Big{]},\\
    n(x>0,k>0,t) =& \sum_{s=0}^{\infty}f_k\,T_k R_k^s \Big{[}\theta(-x-2 s L_S+t\sin[k]) \notag\\
    &- \theta(-x-2 s L_S-L_S+t\sin[k]) \notag \\
    & + \theta(x+2s L_S+2 L_S -t\sin[k]) \notag \\
    &-\theta(x+2s L_S+L_s-t\sin[k])\Big{]}.
    \label{supp_eq:n_Res_theta}
\end{align}

We will now show how Eq.~\eqref{supp_eq:n_sys_theta}-\eqref{supp_eq:n_Res_theta}, reduces to Eq.~(10) and Eq.~(11) of the main text. We first consider Eqs.~\eqref{supp_eq:n_sys_theta} and \eqref{supp_eq:n_Res_theta}.
Using the property of $\theta$ function, $\theta(x)+\theta (-x)=1$, in Eqs.~\eqref{supp_eq:n_sys_theta} and \eqref{supp_eq:n_Res_theta}, it can be easily shown that the above equations boil down to the following forms,
\begin{align}
\label{supp_eq:n_sys_theta_simplified}
    n(x<0,k>0,t) =&\sum_{s=0}^{\infty}f_k\,R_k^s \Big{[}\theta(-x-2 s L_S+t\sin[k]) \notag\\
    & - \theta(-x-2s L_S-2 L_S +t\sin[k])  \Big{]},\\
  n(x>0,k>0,t) =& \sum_{s=0}^{\infty}f_k\,T_k R_k^s \Big{[}\theta(-x-2 s L_S+t\sin[k]) \notag\\
    & - \theta(-x-2s L_S-2 L_S +t\sin[k])\Big{]}.
\label{supp_eq:n_Res_theta_simplified}
\end{align}
Now, for the simplification of Eq.~\eqref{supp_eq:n_sysm_theta}, putting $s' = s+1$ for the first term, we can write,
\begin{align}
\label{supp_eq:n_sysm_theta_simplify1}  n(x<0,k<0,t) \!\!=\!\!& \sum_{s'=1}^{\infty}f_k \,R_k^{s'} \Big{[}\theta(x\!-\!2s' L_S+ 2 L_S\!-\!t\sin[k]) \notag\\
  &-\theta(x-2s' L_S+ L_S-t \sin[k]) \Big{]}\notag \\
    & + \sum_{s=0}^{\infty}f_k R_k^s\Big{[}\theta(-x+2s L_S + t\sin[k])\notag \\
    &-\theta(-x+2s L_S-L_S+t \sin[k]) \Big{]}.
\end{align}
Since the argument inside $\theta(-x+2s L_S-L_S + t \sin[k])$ is always negative for $s=0$ with $k<0$, Eq.~\eqref{supp_eq:n_sysm_theta_simplify1} can be written as,
\begin{align}
\label{supp_eq:n_sysm_theta_simplify2}  n(x<0,k<0,t) \!\!=\!\!& \sum_{s=1}^{\infty}f_k \,R_k^{s} \Big{[}\theta(x-2s L_S+ 2 L_S-t\sin[k]) \notag\\
  &-\theta(x-2s L_S+ L_S-t \sin[k]) \notag \\
  &-\theta(-x+2s L_S-L_S+t \sin[k])
  \Big{]}\notag \\
    & + \sum_{s=0}^{\infty}f_k R_k^s\Big{[}\theta(-x+2s L_S + t\sin[k]) \Big{]}.
\end{align}
where, note that we have changed the dummy index $s'$ back to s in the first summation.
Again using the property of $\theta$ function $\theta(x)+\theta(-x)=1$, Eq.~\eqref{supp_eq:n_sysm_theta_simplify2} can be simplified to,
\begin{align}
\label{supp_eq:n_sysm_theta_simplify3}   
 &n(x<0,k<0,t) = \sum_{s=1}^{\infty}-f_k \,R_k^{s} \theta(-x+2s L_S-2 L_S\notag \\
 &+t\sin[k])+\sum_{s=0}^{\infty}f_k \,R_k^{s} \theta(-x+2s L_S+t\sin[k]).
\end{align}
As the argument inside the theta function for the first term is always negative for $s=0$, we can extend the first summation to $s = 0$ and rewrite Eq.~\eqref{supp_eq:n_sysm_theta_simplify3} as,
\begin{align}
\label{supp_eq:n_sysm_theta_simplify4}   
 n(x<0,k<0,t) =
 &\sum_{s=0}^{\infty}f_k \,R_k^{s} \Big{[}\theta(-x+2s L_S+t\sin[k])- \notag \\
&\theta(-x+2s L_S-2 L_S+t\sin[k]) \Big{]} .
\end{align}
Now, we want to write Eq.~\eqref{supp_eq:n_sysm_theta_simplify4} in the form of Eq.~\eqref{supp_eq:n_sys_theta_simplified}. Using, $x'=-x-2 L_S$ and $k'=-k>0$ (this is like putting a mirror on $x=-L_S$), Eq.~\eqref{supp_eq:n_sysm_theta_simplify4} takes the form,
\begin{align}
\label{supp_eq:n_sysm_theta_simplify5}   
 &n(x'<0,k'>0,t) =
 \sum_{s=0}^{\infty}f_{k'} \,R_{k'}^{s} \Big{[}\theta(x'+2s L_S+ \notag \\
&2 L_S-t\sin[k'])-\theta(x'+2s L_S-t\sin[k']) \Big{]} .
\end{align}
Note that $x'\in[-2L_S,-L_S]$. Again, using the property of the theta function (and changing the dummy variables $x'$ and $k'$ to $x$ and $k$ respectively), we can easily show that Eq.~\eqref{supp_eq:n_sysm_theta_simplify5} takes the following form,
\begin{align}
\label{supp_eq:n_sysm_theta_simplify6}
    n(x<0,k>0,t) =&\sum_{s=0}^{\infty}f_k\,R_k^s \Big{[}\theta(-x-2 s L_S+t\sin[k]) \notag\\
    & - \theta(-x-2s L_S-2 L_S +t\sin[k])  \Big{]}.
\end{align}
Eq.~\eqref{supp_eq:n_sys_theta_simplified} and \eqref{supp_eq:n_sysm_theta_simplify6}, together gives the Eq.~(10) of the main text and Eq.~\eqref{supp_eq:n_Res_theta_simplified} is the Eq.~(11) for the main text.


\section{Exact form of hydrodynamic entropy at $\beta=\infty$ for conformal defect and agreement with exact microscopics}

\begin{figure*}
    \centering
    \includegraphics[width=\linewidth]{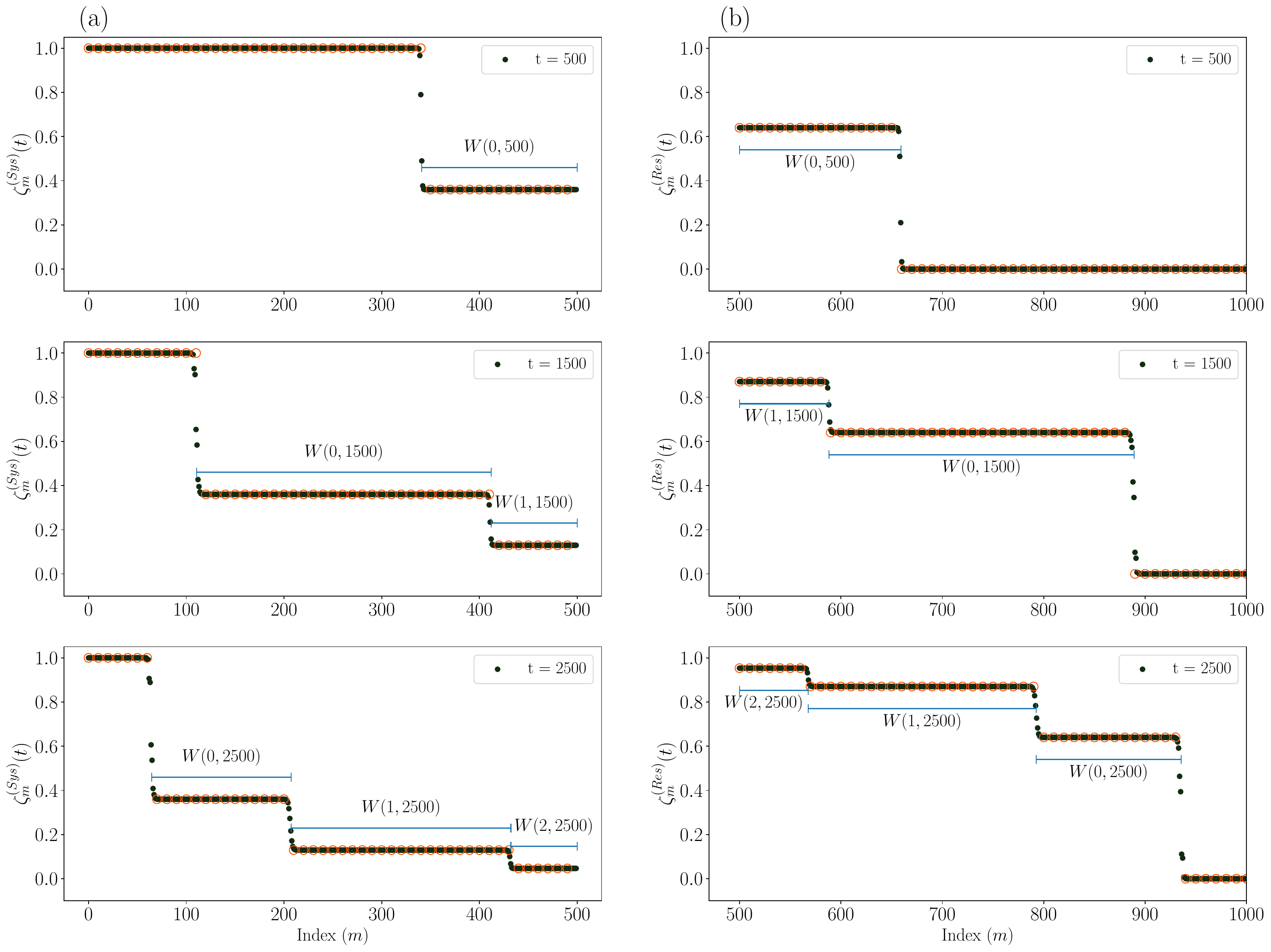}
    \caption{Eigenvalues for (a) the system part of the correlation matrix and (b) the reservoir part of the correlation matrix, for the setup with a conformal defect. The black data points show the numerical results, and the red hollow circles show the analytical results. The width of each plateau region is given by the function $W(s,t)$, which is given in Eq.~\eqref{supp:W_st}. Here $L_S = 500$ and $g_c = 0.4$}
    \label{fig:Conf_Wst}
\end{figure*}

\begin{figure}
    \centering
    \includegraphics[width=\linewidth]{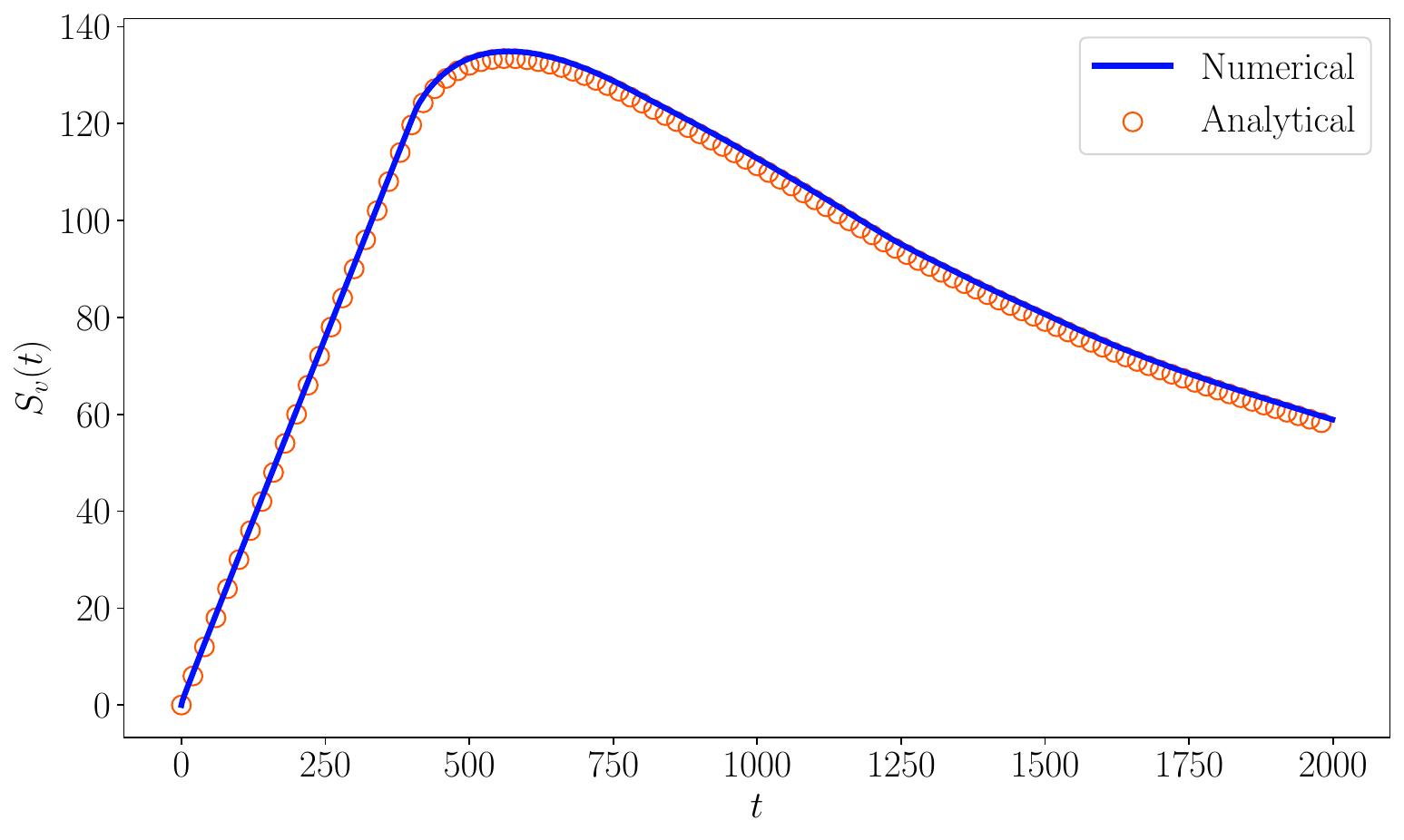}
    \caption{$S_{v}$ as a function of time for the setup with a conformal defect. The blue solid line depicts the numerically obtained von Neumann entropy, and the hollow red circles depict the analytical result obtained by plotting Eq.~\eqref{supp:hydro_conf}. Here $L_S = 200$ and $g_c = 0.4$}
    \label{fig:S_vN_conform}
\end{figure}

With $\beta=\infty$ for a setup with a conformal defect,  with the phase space density given by Eq.~\eqref{supp_eq:Hrdro_sol} and using the hydrodynamic entropy formula from the End Matter (Eq.~(EM5) and Eq.~(EM6)), it can be shown that the system hydrodynamic entropy has the form,
\begin{align}
\label{supp:hydro_conf}
    S_{\textrm{Hydro}}^{\textrm{Sys},\beta = \infty}(t) &= \sum_{s = 0}^{\infty} W(s,t) \Big{[}(1-\lambda^2)^{s+1} \log[(1-\lambda^2)^{s+1}]\notag\\
    & + (1-(1-\lambda^2)^{s+1})\log[(1-(1-\lambda^2)^{s+1})]\Big{]}
\end{align}
where recall that $(1-\lambda^2) = R_k$ i.e. the reflection probability in the presence of a conformal defect. Here, $s$ is the number of reflections that occur at the defect. $W(s,t)$ can be explicitly determined for a conformal defect by integrating Eq.~EM5 (End Matter) over $x$ and $k$. For instance, for $t<2L_S$, it can be shown that,
\begin{equation}
    W(s,t<2L_s) = \begin{cases}
        \frac{t}{\pi} \hspace{3em} s = 0\\
        0 \hspace{3.5em} s>0
    \end{cases}
\end{equation}
This is consistent with Ref.~\cite{SKD2024}. Similarly, for $2L_S<t<4L_S$, $W(s,t)$ takes the form,
\begin{equation}
    W(s,t) = \begin{cases}
        \frac{1}{\pi}\Big[t + L_S\pi - \sqrt{t^2-4L_S^2} \\
        - 2L_s\sin^{-1}[\frac{2L_S}{t}] \Big]\hspace{8em}s = 0\\\\
        \frac{1}{\pi}\Big[-\pi L_S + \sqrt{t^2-4L_S^2}+\\
        2L_S\sin^{-1}[\frac{2L_S}{t}]\Big] \hspace{8.5em} s=1\\\\
        0 \hspace{15em} s>1.
    \end{cases}
\end{equation}

 \begin{figure}[ht]
\centering
\includegraphics[width=1.0\columnwidth]{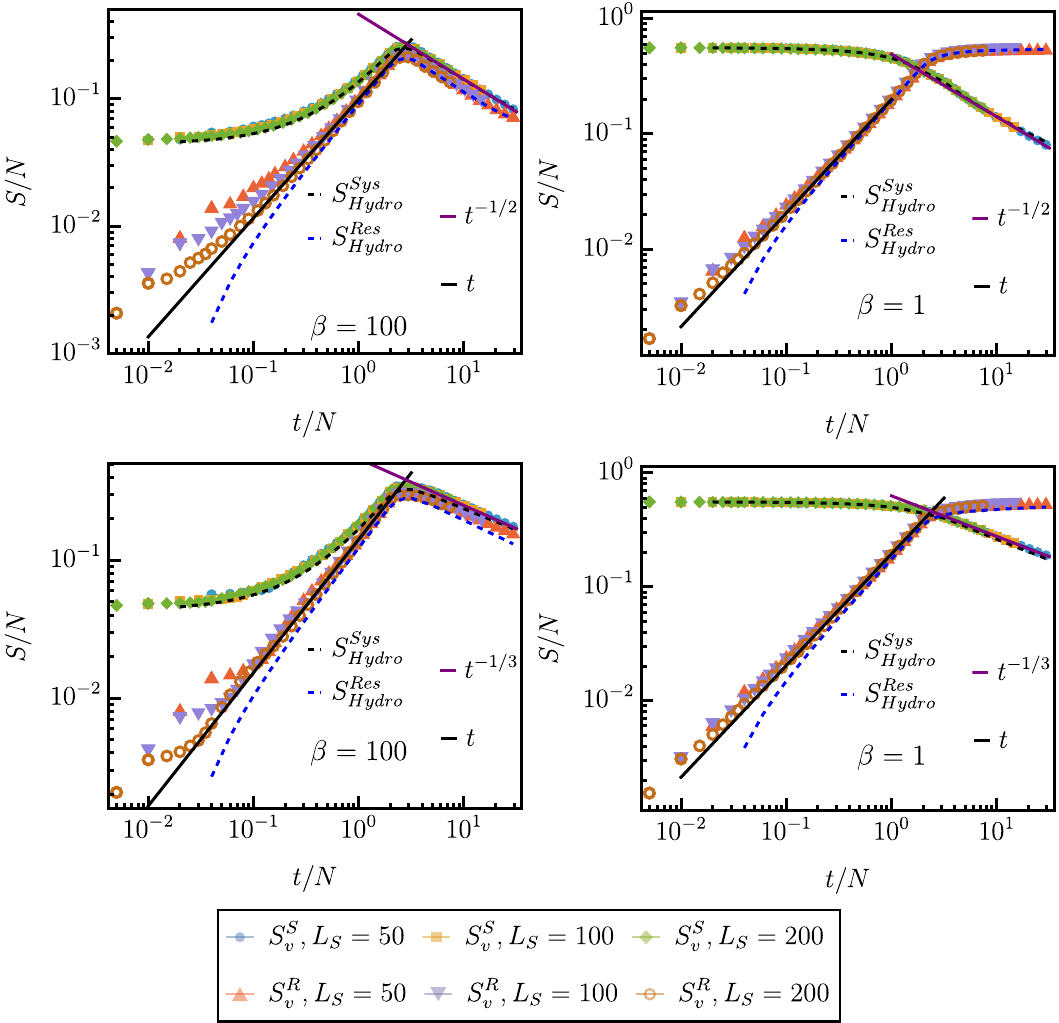} 
\caption {The figure shows various entropies for different values of $\beta$ for the hopping and density defect. Silimarly like the conformal defect, the mixed state von Neumann entropy of the system and the bath are in excellent agreement with the corresponding finite temperature mixed state hydrodynamic entropies. The figure also clearly demonstrates the need to have modified hydrodynamic formalism for the reservoir entropy. A naive hydrodynamic entropy in the reservoir (blue-dashed line) would predict a monotonic growth very different from that of the 
mixed state von Neumann entropy in the bath. } 
\label{supp_mixed_other_defect} 
\end{figure}

Generalizing it to any time $t$, it is easy to see that for $t<mL_S$, where $m$ is an integer greater than or equal to 0, the contribution comes from $s = 0,\cdots,m$. The $W(s,t)$ takes the form,
\begin{widetext}
\begin{align}
    W(s,t) = \begin{cases}
        \displaystyle 
\frac{1}{\pi}(
-\pi s L_S 
+ \sqrt{t^{2} - (2sL_S)^{2}}
+ 2sL_S \sin^{-1}\!\left(\frac{2sL_S}{t}\right)
)\hspace{8.5em}\textrm{for}\,\,\,\,\,\,\,\,\,\,\,\, 2sL_S<t < 2(s+1)L_S\\\\
        \displaystyle
\frac{1}{\pi}((s+2)\pi L_S - 2\sqrt{t^{2} - 4(s+1)^{2}L_S^{2}}
+ \sqrt{t^{2} - (2sL_S)^{2}}\\
+ (2sL_S)\sin^{-1}\!\left(\frac{2sL_S}{t}\right)
- 4(s+1)L_S\sin^{-1}\!\left(\frac{2(s+1)L_S}{t}\right)
) \hspace{7.5em}\textrm{for}\,\,\, 2(s+1)L_S <t < 2(s+2)L_S\\\\
        \displaystyle
\frac{1}{\pi}(\sqrt{t^{2} - 4(s+2)^{2}L_S^{2}}- 2\sqrt{t^{2} - 4(s+1)^{2}L_S^{2}}
+ \sqrt{t^{2} - 4s^{2}L_S^{2}}\\
+ 2sL_S \sin^{-1}\!\left(\frac{2sL_S}{t}\right)
- 4(s+1)L_S\sin^{-1}\!\left(\frac{2(s+1)L_S}{t}\right)
+ 2(s+2)L_S\sin^{-1}\!\left(\frac{2(s+2)L_S}{t}\right)
) \hspace{1.5em}\textrm{for}\,\,\, t>2(s+2)L_S .
    \end{cases}
    \label{supp:W_st}
\end{align}
\end{widetext}
Upon inspection of the eigenvalues of the correlation matrix for $\beta = \infty$ case (described in the main manuscript),  $\lceil W(s,t)\rceil$, given Eq.~\eqref{supp:W_st}, can be identified as the number of eigenvalues at the saturation level $R_k^{s+1}$. Thus, the von Neumann entropy given in Eq.~\eqref{supp:SvN} can be written as,
\begin{align}
\label{supp:SvN_W}
    S^{\beta = \infty}_{v}(t) &\approx \sum_{s} W(s,t) \Big{[}(1-\lambda^2)^{s+1} \log[(1-\lambda^2)^{s+1}]\notag\\
    & + (1-(1-\lambda^2)^{s+1})\log[(1-(1-\lambda^2)^{s+1})]\Big{]}
\end{align}
It can be easily seen that Eq.~\eqref{supp:hydro_conf} and Eq.~\eqref{supp:SvN_W} are approximately equivalent.
Note that Eq.~\eqref{supp:SvN_W} is correct only upto the leading order, as there will be a sub-extensive number of eigenvalues with does not lie on the constant saturation levels, which will give logarithmic corrections as discussed in the main text.

\section{Other defects: density defect, hopping defect}

In the main text, we discussed mainly about the conformal defect. However, numerically we have observed that hydrodynamic entropy matches well with the exact von-Neumann entropy for all times for other defects such as the hopping defect and the density defect. For the hopping defect and density defect, the coupling Hamiltonian takes the following form:
\begin{align}
 \hat{V} = 
\begin{cases}
-g[\hat{c}^\dagger_{0}\hat{c}_{1} + \hat{c}^\dagger_{1}\hat{c}_0] +g_c [\hat{c}^\dagger_{0}\hat{c}_{0} + \hat{c}^\dagger_{1}\hat{c}_1] & \!\!\!\!\!\text{(density defect)}  \\
-g_c[\hat{c}^\dagger_{0}\hat{c}_{1} + \hat{c}^\dagger_{1}\hat{c}_0] &  \!\!\!\!\!\!\!\!\!\!\!\text{(hopping defect)}.
\end{cases}
\end{align}
In Fig.~\ref{supp_mixed_other_defect}, we have shown that for different values of $\beta$, even for hopping and density defects, the hydrodynamic entropy for the system and the reservoir (with the modified description of phase space density) agrees with the von-Neumann entropy calculated from the exact microscopics. It can be seen that the reservoir hydrodynamic entropy scales as $t$ for all types of defect at any value of $\beta$. 
It can be seen that at $\beta  = 0$ the reservoir hydrodynamic entropy saturates at a constant value, but as $\beta$ is increased, the saturation value decreases. As $\beta\xrightarrow{}\infty$, the reservoir hydrodynamic entropy decays to zero at late times and shows a Page curve. The system's hydrodynamic entropy initially has a finite value for $\beta = 0$, and it decays to zero at late times. The decay of system hydrodynamic entropy shows a defect-dependent scaling. For a hopping defect, both the system hydrodynamic entropy and exact von-Neumann entropy decays as $t^{-1/2}$ (however, late time calculations from hydrodynamics show that it should decay as $t^{-1/3}$ even for hopping defect \cite{SKD2024} at $\beta=\infty$), and for a density defect, it scales as $t^{-1/3}$. As $\beta$ is increased, the initial system hydrodynamic entropy decreases. It begins at zero for $\beta\xrightarrow{}\infty$ and it also shows a Page curve which exactly coincides with that of the reservoir hydrodynamic entropy.

Similar to the conformal defects, for the above discussed defects, we can provide the hydrodynamic description of full counting statistics.


\end{document}